\documentclass[12pt]{iopart}
\usepackage{graphicx}
\usepackage[usenames]{color}

\def \ds {d_\mathrm{s}}
\def \di {d_\mathrm{i}}
\def \Tc {T_\mathrm{c}}
\def \Jc {J_\mathrm{c}}
\def \ax {a_x}

\begin{document}

\title{Flux Avalanches in Nb Superconducting Shifted Strip Arrays}
\author{Y.~Tsuchiya$^1$, Y.~Mawatari$^2$, J.~Ibuka$^1$, S.~Tada$^1$, S.~Pyon$^1$, S.~Nagasawa$^3$, M.~Hidaka$^3$, M. Maezawa$^2$, and T.~Tamegai$^1$}
\address{$^1$Department of Applied Physics, The University of Tokyo, Hongo, Bunkyo-ku, Tokyo 113-8656, Japan}
\address{$^2$National Institute of Advanced Industrial Science and Technology (AIST), Tsukuba, Ibaraki 305-8568, Japan}
\address{$^3$Superconductivity Research Laboratory, International Superconductivity Technology Center (ISTEC), Tsukuba, Ibaraki 305-8501, Japan}

\eads{0214689544@mail.ecc.u-tokyo.ac.jp}
\begin{abstract}
Flux penetrations into three-dimensional Nb superconducting strip arrays, where two layers of strip array are stacked by shifting a half period, are studied by using magneto-optical imaging method.
Flux avalanches are observed when the overlap between the top and bottom layers is large even if the width of each strip is well below the threshold value.
In addition, anomalous linear avalanches perpendicular to the strip are observed in the shifted strip array when the overlap is very large and the thickness of superconductor is greater than the penetration depth.
We discuss possible origins of the flux avalanches including the linear ones by considering the flux penetration calculated by the Campbell method assuming the Bean model.
\end{abstract}
\pacs{74.78.Fk, 68.60.Dv, 74.25.Ha, 74.25.Op}

\submitto{\SUST}
\maketitle

\section{Introduction}

Superconductors have been fabricated into three-dimensional structures to realize their unique functionalities.
They can be used as metamaterials with unusual electric permittivity or magnetic permeability~\cite{Pendry:2006hq,Schurig:2006ws},
which can realize the magnetic cloaking without disturbing the external magnetic lines of force compared with the conventional magnetic shielding by superconductors or mu-metals~\cite{Magnus:2008fw,Narayana:2011ht,Gomory:2012bo}.
The rapid single-flux-quantum device is also a major application which can replace the conventional CMOS semiconductor device for its high operation speed with low power consumption~\cite{Chen:1999uv}.
Additionally, the superconducting Roebel cable with a three-dimensional structure can suppress AC losses by twisting superconducting tapes~\cite{Goldacker:2006fc}.
For those applications, bulk magnetization measurements, evaluations of device operation, or transport measurements have been extensively undertaken as well as theoretical calculations~\cite{Nii:2012gf}.
However, little is known about the local magnetic response of such three-dimensional superconductors.
Therefore, it is of great importance to investigate their local magnetic properties to obtain scientific and technological information on the three-dimensional superconductors and to improve the performance of their applications.

The magnetic flux penetrates into superconductors as quantized vortices under magnetic fields above the lower critical field,
and the vortices arrange themselves in a self-organized manner into the critical state,
where the Lorentz force by the shielding current balances the pinning force by defects~\cite{Bean:1962vg}.
On the other hand, unique phenomena known as flux avalanches (flux jumps) due to thermo-magnetic instabilities are often observed when the heat generated by the vortex motion surmounts the heat diffusion~\cite{Swartz:1968dk,Duran:1992wu}.
This instability becomes an obstacle for the applications of superconductors due to quenching~\cite{Bellis:1994vb} or generation of noise~\cite{Olson:1997tk}.
The condition for flux avalanches depends on various kinds of parameters; temperature, applied field, thermal conductivity, the geometry of samples, and so on~\cite{Denisov:2006cs}.
Flux avalanches can occur at very low magnetic fields in thin film samples due to stronger demagnetization effect~\cite{Colauto:2009gf}.
This effect can be enhanced even more in superconducting structures consisting of many thin films arranged close by~\cite{Mawatari:1996we}.
Recently, electromagnetic responses in hexagonal arrays of superconducting strips have been reported theoretically as a metamaterial with stronger anisotropy in magnetic permeability~\cite{Mawatari:2012hj}.
The larger magnetic anisotropy in the system is suitable to realize magnetic cloaking and other applications by controlling magnetic field distribution.
However, it always accompanies a risk of thermomagnetic instability due to stronger demagnetizing effect.
Therefore, it is important to investigate conditions for flux avalanches and get information on the controlling parameter to such phenomena.

In this paper, we extract two layers from the hexagonal array of superconducting strips,
and investigate its local magnetic response with changing temperatures, fields, and the sizes of the samples.
Hereafter, we call these three-dimensional structures as ``shifted strip arrays (SSAs)''.

\section{Experiments}

\begin{figure}[htbp]
    \centering
    \includegraphics[width=8cm]{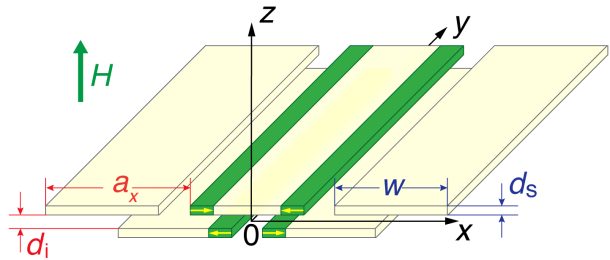}
    \caption{Schematic drawing of shifted strip array.
    Each strip has width $w$ and thickness $\ds$.
    In each layer, strips are arranged periodically along $x$ axis with a period of $\ax$
    and separated into two layers with insulating gap $\di$.
    Schematic flux penetration into strips are shown with green regions in the middle unit cell,
    which propagates from the edges to the center (yellow arrows).}
\end{figure}

\begin{table}[htbp]
\caption{Name list of shifted strip array samples.
 Each sample has thickness of superconducting film $\ds$ and thickness of insulator gap $\di$.}
\begin{indented}
\item[]\begin{tabular}{@{}lll}
\br
Name&$\ds$ (nm)&$\di$ (nm)\\
\mr
SSA50-50&50&50\\
SSA100-100&100&100\\
SSA200-200&200&200\\
SSA300-300&300&300\\
SSA300-50&300&50\\
SSA500-50&500&50\\
\br
\end{tabular}
\end{indented}
\end{table}

We have fabricated several kinds of SSAs of Nb ($\Tc = 9.2$~K) on Si substrates by using DC magnetron sputtering, photolithography, SF$_6$ reactive ion etching, caldera planarization technique~\cite{Nagasawa:2004de,Satoh:2005fs}, and have cut unnecessary parts by using focused-ion-beam etching technique.
Figure 1 shows the schematic drawing of SSA, which corresponds to two layers of hexagonal strip arrays.
The array consists of strips with a fixed width of $w = 8~\mu$m and thicknesses $\ds$ ranging from 50 to 500~nm.
Strips are arranged periodically along $x$ axis with period $\ax$ ranging from 9 to 16~$\mu$m.
Two layers of such arrays are displaced by $\ax$/2 along $x$-axis and stacked by sandwiching SiO$_2$ insulating layer with thicknesses $\di$ ranging from 50 to 300~nm.
We define names of SSAs as SSA$\ds$-$\di$ as shown in Tab. 1.

Flux distributions in SSAs are observed under the external magnetic field $H$ ($0-40$~mT) along $z$ axis after cooling them down to temperature $T$ ranging from 4.5 to 8.5~K under zero-field.
To visualize flux penetrations in SSAs, we use the magneto-optical (MO) imaging technique in which the spatial variation of the out-of-plane flux density is detected by using the Faraday effect in a ferromagnetic garnet film~\cite{Polyanskii:1989vl,Dorosinskii:1992tp}.
The gap between the sample and the garnet film has been minimized by gently pushing down the garnet film with two small phosphor bronze plates.
In MO measurements, we used a commercial optical microscope (OLYMPUS BX30MF) with a x20 objective lens (Nikon CF Plan 20X) and a cooled-CCD camera with 12-bit resolution (ORCA ER, Hamamatsu).
The samples are cooled down to temperatures ranging from 4.2 K to 8.5 K under zero-field by using a He-flow cryostat (MicrostatHighRes II, Oxford Instruments).
Magnetic field along $z$-axis was applied isothermally at each temperature.
The obtained MO images are integrated over tens of images and subtracted by a zero-field background image to improve the magnetic resolution and to suppress artifacts originated from in-plane magnetic domains or scratches on the garnet film~\cite{Soibel:2000vf}.
It should be noted that even single vortices can be observed by this technique~\cite{Soibel:2000vf,Goa:2001vz,Yasugaki:2002hk,Tsuchiya:2010fi}.

\section{Results and Discussion}

\begin{figure}[htbp]
    \centering
    \includegraphics[width=8cm]{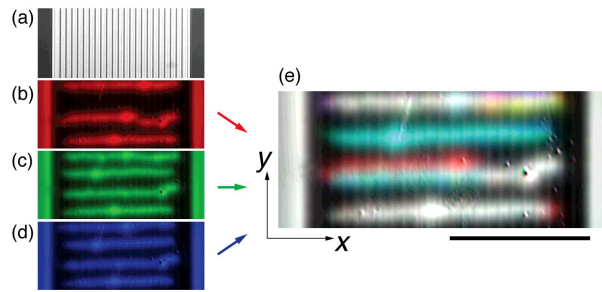}
    \caption{(a) Optical micrograph of SSA300-50 with $\ax = 9~\mu$m.
    (b)-(d) Three MO images of flux penetrations after cooling down to 4.2~K and applying $H = 10$~mT.
    (e) Image obtained by adding the three MO images.
    Black bar shows 100~$\mu$m length scale.}
\end{figure}

Figure 2(a) shows an optical image of SSA300-50 with $\ax = 9~\mu$m.
Numbers of strips in $x$ direction are chosen to limit the samples within a range of 200~$\mu$m.
In this sample, 23 and 24 strips are fabricated on the top and bottom layers, respectively.
Lengths of strips in $y$ direction are also 200~$\mu$m, which is long enough to consider the system as a collection of infinite strips.
MO images of flux penetrations into the sample are shown in Figs.~2(b)-2(d) under the same condition at 4.2~K and under $H = 10$ mT repeatedly.
These images are colored in three kinds of channels (red, green, and blue) and are added into an image as shown in Fig.~2(e).
In this combined image, the gray scale regions show reproduced flux penetrations, known as the usual critical state.
On the other hand, colored areas indicate that the flux front is not reproducible.
This phenomenon is known as the flux avalanche caused by the thermo-magnetic instability, which rapidly appears with increasing field and has non-reproducible shapes~\cite{Altshuler:2004vt}.
Some avalanches, however, repeatedly occur at similar locations possibly due to defects in the superconductor or the stray field generated by the magnetic domains in the indicator film.
Flux avalanches usually occur with their positions far from each other due to the nonlocal repulsive force between flux in thin films,
and the observed flux avalanches similarly do not overlap but align.
There are two noteworthy points different in flux avalanches in SSA different from those in plain films.
One is about morphology of flux avalanches.
The observed flux avalanches have linear shapes traversing many strips.
Flux avalanches in plain films are known to have various kinds of shapes~\cite{Vestgarden:2013is}: uniform~\cite{Prozorov:2006jh}, finger-like~\cite{Altshuler:2004ww}, and dendritic~\cite{Denisov:2006ho}.
On the contrary, flux avalanches in SSAs likely jump from one strip to another and tend to form anomalous linear shapes perpendicular to the strips as shown in Figs.~2(b)-2(d).
Another point is about the starting point of flux avalanches.
They do not start at the outermost edges of SSAs as they do in plain films.
In addition, we should not overlook that linear flux avalanches have knots around the middle of the lines and they may have initiated there.
However, the time-resolved measurements of flux propagation is needed to discuss where flux avalanches start from~\cite{Leiderer:1993us}.

\begin{figure}[htbp]
    \centering
    \includegraphics[width=8cm]{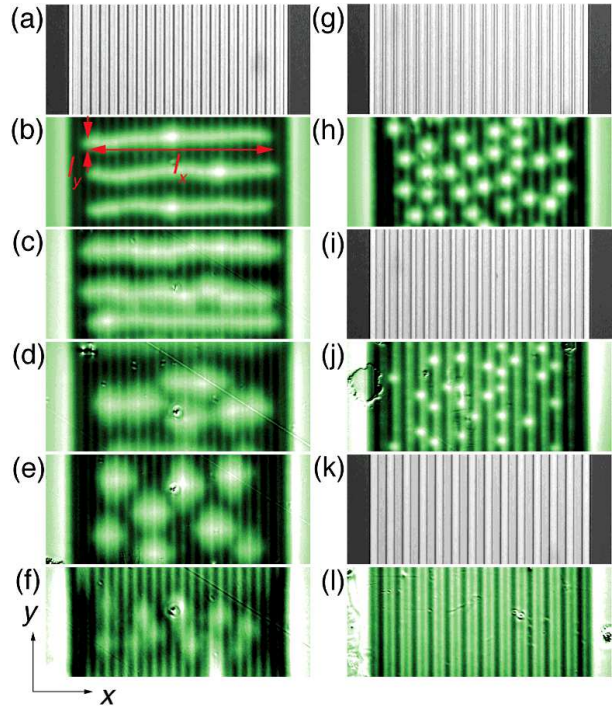}
    \caption{Temperature and $\ax$ dependences of flux penetrations in SSA300-50.
    (a) An optical micrograph and MO images of flux penetrations with $\ax = 9~\mu$m at (b) 4.2 K and 10 mT, (c) 6 K and 9 mT, (d) 7 K and 6 mT, (e) 8 K and 4 mT, and (f) 8.5 K and 2.2 mT.
    The sizes of flux avalanches along $x$ and $y$ direction are defined as $l_x$ and $l_y$, respectively.
    Optical micrographs and MO images at 4.2 K with $\ax =$ (g), (h) 11~$\mu$m at 11 mT, (i), (j) 13~$\mu$m at 8 mT, and (k), (l) 16~$\mu$m at 8 mT.}
\end{figure}

Figure 3 shows $T$ and $\ax$ dependences of flux penetrations in SSA300-50.
The MO images shows flux distribution after applying magnetic fields isothermally at each temperatures.
With $\ax = 9~\mu$m, flux avalanches occur at high temperatures up to 8.5~K (0.92$\Tc$) as shown in Fig.~3(f).
Generally, flux avalanche in strips or films occur below a characteristic field-dependent temperature, so-called threshold temperature, and in range of magnetic fields above a threshold field and below a certain upper field in isothermal conditions~\cite{Denisov:2006cs,Yurchenko:2007fv,Colauto:2008db}.
The typical value of the threshold temperature in Nb film is about 5 K~\cite{Denisov:2006ho,Colauto:2008db,Nowak:1997uc,Behnia:2000uk}.
According to Eq. (1) of Ref.~\cite{Denisov:2006ho}, flux avalanches occur only below threshold temperature $\sim$ 1~K in an isolated superconducting single strip with thickness of 300~nm and width of 8~$\mu$m assuming typical physical parameters of Nb~\cite{Denisov:2006ho}.
Therefore, it seems reasonable to suppose that the three-dimensional structures of SSAs enhance flux avalanches.
We will discuss how the structures induce flux avalanches in latter part of this paper.
In the followings, detailed account of linear flux avalanches is given.
As shown in Fig.~3(b), the linear avalanches at 4.2~K expand over the entire width of the sample (200~$\mu$m).
At higher temperatures (Figs.~3(c)-3(f)), the length of the flux avalanche along $x$ axis, $l_x$, becomes smaller and avalanches form small fragments.
It is obvious that the average length $l_x$ monotonically decreases with increasing $T$.
In contrast, the behavior of the width of the linear flux avalanche along $y$ axis, $l_y$, is different.
It is the smallest at 4.2~K (Fig.~3(b)), and increases with increasing $T$ as shown in Figs.~3(c)-3(e), followed by a decrease near $\Tc$ (Fig.~3(f)).
According to earlier theory and experiments \cite{Denisov:2006cs,Johansen:2002vs}, sizes of flux avalanches both along $x$ and $y$ directions increase monotonically with increasing $T$.
It is opposite to out result that the length $l_x$ decreases as temperature increases.
There is no satisfactory explanation about this anomalous temperature dependence at the present stage.
On the other hand, the positive temperature dependence of $l_y$ in SSAs below 8~K agrees with the theory for the single strip.
In addition, the anomaly of $l_y$ at 8.5~K (Fig.~3(f)) can be explained by the large penetration depth near $\Tc$ which is even larger than the thickness $\ds$ when we assume two-fluid model for its $T$ dependence~\cite{Gubin:2005id}.
Next, $\ax$ dependence of the flux avalanches at 4.2~K is shown in Figs.~3(g)-3(l).
With increasing $\ax$, the linear avalanches disappear at 11~$\mu$m as shown in Fig.~3(h).
Instead of the linear flux avalanches, spot-like avalanches appear in each strip.
With larger $\ax$, the size of spots shrinks (Fig.~3(j)).
Finally, no flux avalanches occur in SSA with $\ax = 16~\mu$m (Fig.~3(l)), confirming that the threshold temperature in single strips with $w = 8~\mu$m is below 4.2~K as estimated above, because SSA without an overlap can be treated as a quasi two-dimensional array of single strips.
It indicates that the three-dimensional structure in SSA promotes flux avalanches.
In the following section, we discuss possible origins of the linear flux avalanches.

There are two scenarios to explain the origin of the linear avalanches; thermal and magnetic couplings.
The interlayer thermal coupling promotes the linear flux avalanches.
The heat generated by an avalanche diffuses into the surroundings which are usually considered as a heat bath at a fixed temperature.
The heat transfer coefficient is an important parameter to determine the condition for flux avalanches~\cite{Denisov:2006cs}.
However, special care needs to be taken in SSAs because the strips are thermally coupled to each other via thin insulating layers.
Therefore the heat produced by the first flux motion in one layer triggers the motion of another flux in the neighboring layer due to the interlayer thermal coupling.
Besides, the linear flux avalanches can occur due to the magnetic coupling between vortices in two overlapping layers.
The motion of vortices in the bilayer superconductors has been well-studied for its application of the dc superconducting transformer~\cite{Sherrill:1975wa}.
The stray field generated by a vortex in one superconducting layer attracts another vortex in the neighboring layer.
Even when the Lorentz force is applied only on one of the vortices, the other vortex is dragged by the magnetic coupling force.
Of course, the two couplings are not mutually exclusive rather they can cooperate to promote the linear flux avalanche.

\begin{figure}[htbp]
    \centering
    \includegraphics[width=8cm]{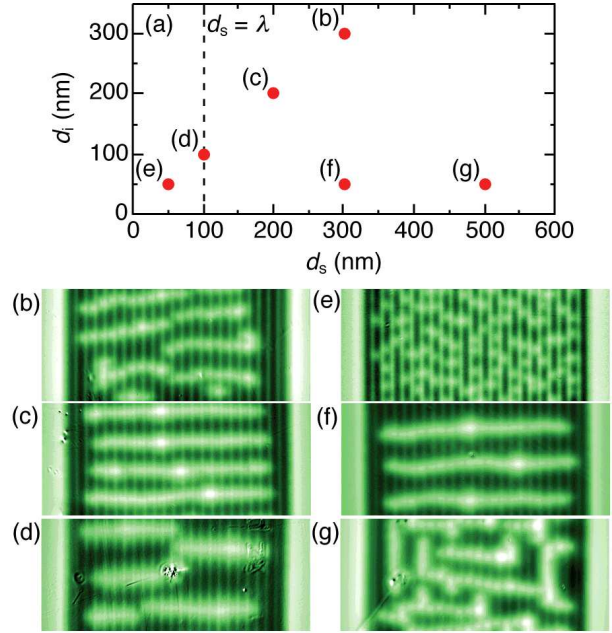}
    \caption{(a) Schematic diagram of $\di$ and $\ds$ for the SSA samples.
    MO images of flux avalanches in SSAs with different $\ds$ and $\di$ with $\ax = 9~\mu$m at 4.2~K under $H=$ (b) 11 mT, (c) 9 mT,(d) 5.5 mT,(e) 5 mT,(f) 10 mT, (g) 40 mT.
}

\end{figure}

Figure 4 shows MO images for SSAs with different $\di$ and $\ds$ with $\ax = 9~\mu$m at 4.2~K.
Considering $\di$ dependence with a fixed $\ds$ (see Figs.~4(b) and 4(f)), the length $l_x$ becomes larger with a smaller $\di$.
This result is consistent with thermal/magnetic coupling scenarios.
Next, $\ds$ dependence with a fixed $\di$ is non-monotonic as shown in Figs.~4(e)-4(g), indicating that it is difficult to explain the $\ds$ dependence only by the above scenario because the thermal/magnetic coupling should hold constant with the same $\di$.
It is especially anomalous that no linear avalanche is observable with small $\ds$ as shown in Fig.~4(e).
To explain this behavior, a dashed line for the condition with $\ds \sim \lambda = 102$~nm at 4.2~K is plotted in Fig.~4(a), where $\lambda$ is the penetration length~\cite{Gubin:2005id}.
As noted in the temperature dependence, we confirm the anomaly occurs when $\lambda$ becomes comparable to or larger than $\ds$ (Fig.~3(f)).
Finally, flux avalanches with equal $\ds$ and $\di$ are shown in Figs.~4(b)-4(e).
This dependence is complicated because it has both $\ds$ and $\di$ dependences.
Similar to the $\di$ dependence with fixed $\ds$, the length $l_x$ decreases with increasing $\ds$ when $\ds$ is larger than 200~nm (Figs.~4(b) and (c)).
However, even with thinner $\ds$, the linear avalanches disappear when $\ds$ is less than $\lambda = 102$~nm (Figs.~4(d) and 4(e)).
From the results on $\ax$, $\di$, $\ds$, and $T$ dependences,
the conditions for the linear avalanche can be summarized as follows;
(1) large overlap between top and bottom layers,
(2) thinner $\di$ for the stronger interlayer thermal/magnetic coupling,
(3) thinner $\ds$ for the longer length $l_x$,
and (4) the thickness $\ds$ should be larger than $\lambda$.
In the following section, we discuss the origin of the flux avalanches in SSA by considering the flux distribution.

\begin{figure}[htbp]
    \centering
    \includegraphics[width=16cm]{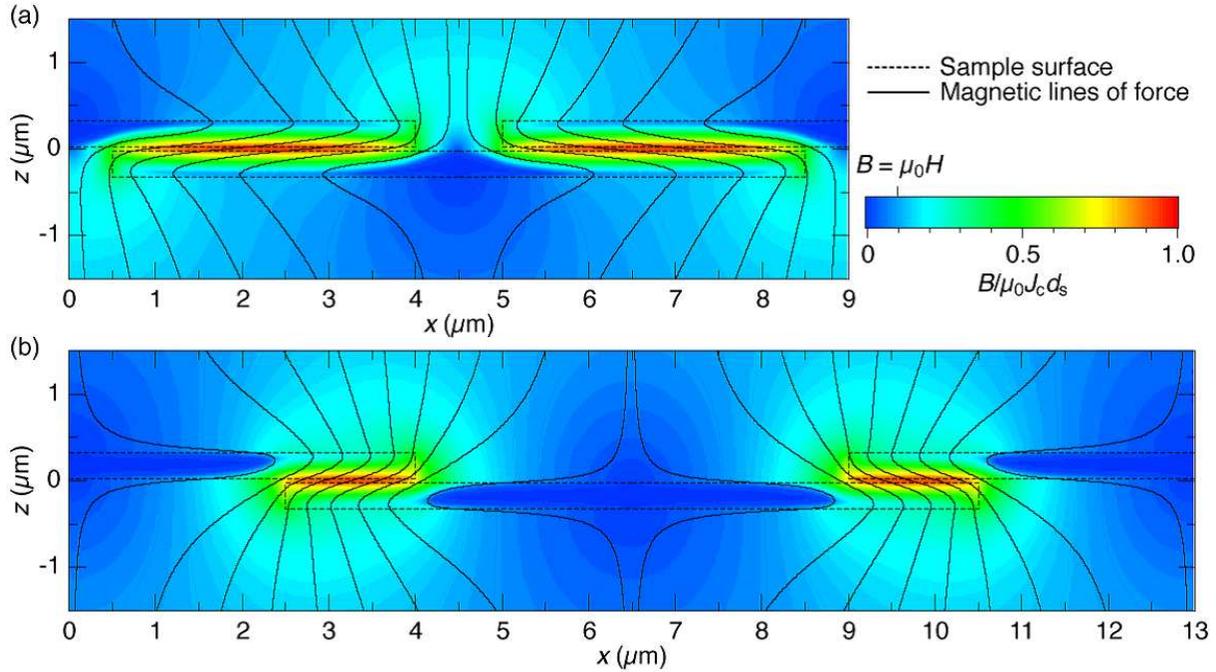}
    \caption{Magnetic lines of force and distributions of magnitude of local flux density in SSA300-50 with $\ax =$ (a) 9~$\mu$m and (b) 13~$\mu$m with assuming the Bean model.}
\end{figure}

The electromagnetic responses of SSAs are calculated by using the Campbell method with assuming the Bean model~\cite{Campbell:2007jj}.
Figure 5 shows the flux distribution in SSA300-50 with $\ax$ = (a) 9 $\mu$m and (b) 13 $\mu$m under applied field of $H = 0.1\Jc\ds$, where $\Jc$ is the critical current density.
The magnitude of the flux density, $B$, is plotted in a color scale.
It is noted that, $B$ becomes 10 times larger than $\mu_0H$ within the overlaps,
and the maximum values are almost the same for different $\ax$'s.
We speculate that this enhancement of the local flux density and its temporal evolution generates the large electric field and heat dissipation when the flux penetrates into SSA,
which finally promotes the flux avalanches at such high temperatures close to $\Tc$.
In addition to $B$ distribution, the magnetic lines of force are shown as black solid lines in Fig.~5.
The lines lie in-plane, indicating that the magnetic coupling is weak in this system compared with the conventional superconducting dc transformer, where flux lines penetrate two films perpendicularly.
Therefore, we can conclude that the most possible origin of linear flux avalanches is the thermal coupling between overlapping strips.
It should be noticed that these numerical calculations corresponds to flux distributions at low $T$ or thick $\ds$ since $\lambda$ is considered to be much smaller than $\ds$. 
Therefore, further discussions on the condition where $\lambda$ becomes comparable to $\ds$ will be required to explain the experimental anomaly at high temperatures.

\begin{figure}[htbp]
    \centering
    \includegraphics[width=8cm]{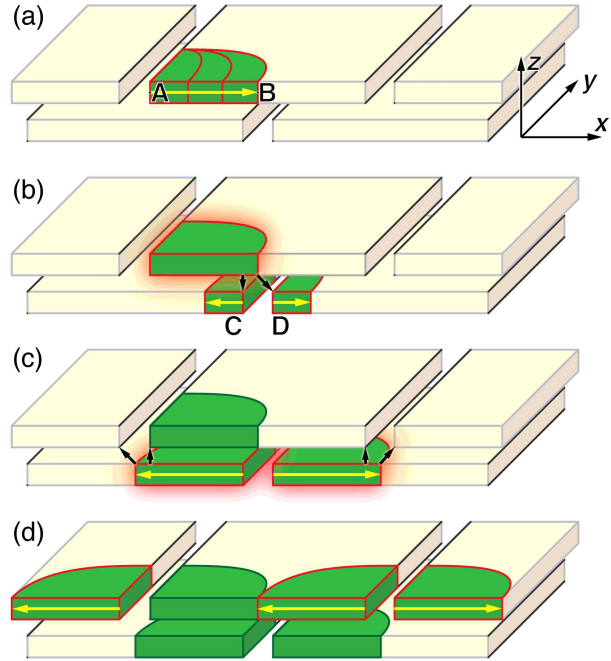}
    \caption{Schematic drawing of the process of a linear flux avalanche.
    (a) A flux avalanche occurs at A in one layer.
    Flux penetration proceeds toward B shown as a yellow arrow.
    (b) Heat dissipation generated by the first avalanche diffuses into surroundings which triggers new avalanches at C and D in another layer.
    Heat transfer is shown as black arrows.
    (c, d) Linear flux avalanches are formed by repeating the same processes.}
\end{figure}

Finally, we explain how interlayer thermal couplings promote linear flux avalanches.
Figure 6 shows a schematic model how a linear avalanche evolves in time.
We assume that the flux front of an avalanche can reach only the middle of the strip and that the thermal diffusion is slower than the magnetic diffusion.
(a) An avalanche occurs at A due to thermal or magnetic fluctuations in one layer and it proceeds towards B in the strip as shown as a yellow arrow.
(b) The heat generated by the avalanche diffuses out into the surroundings as shown by a red shadow.
Then, new avalanches are triggered at C and D due to the thermal fluctuations
which are enhanced by the heat transfers from the original strip via an insulating layer as shown by black arrows.
(c) Linear flux avalanches proceed in the bottom layer by the same processes and the generated heat is transferred to the top layer as shown by black arrows.
(d) New linear avalanches are formed in the top layer as shown by yellow arrows.
However, the processes will stop when the following condition is satisfied.
The heat transfer from on layer to the other is not enough with large $\ax$ or thick $\di$, which is shown by black arrows in Fig.~6(b).
This consideration is consistent with the experimental results that linear avalanches occur with small $\ax$, small $\di$  as shown in Fig.~3 and Fig.~4.

\section{Summary}

Flux penetrations into three-dimensional double-layered superconducting strips, shifted strip arrays, are studied by using magneto-optical imaging method.
Flux avalanches are observed when the period of a shifted strip array is small, or the overlap between the top and bottom layers is large.
Anomalous linear flux avalanches are observed in the shifted strip array when the overlap is very large when the thickness of superconductor is larger than the penetration depth, and when the interlayer gap is small.
We conclude that the linear flux avalanches are caused by the interlayer thermal coupling.
We believe that our work opens a new direction of the study of three-dimensional nanostructured superconductors.

\ack
We thank S. Ooi for his help in the fabrication of shifted strip arrays.

\section*{References}
\bibliography{biblio}

\providecommand{\newblock}{}
\begin{thebibliography}{10}
\expandafter\ifx\csname url\endcsname\relax
  \def\url#1{{\tt #1}}\fi
\expandafter\ifx\csname urlprefix\endcsname\relax\def\urlprefix{URL }\fi
\providecommand{\eprint}[2][]{\url{#2}}

\bibitem{Pendry:2006hq}
Pendry J~B 2006 {\em Science\/} {\bf 312} 1780--1782

\bibitem{Schurig:2006ws}
Schurig D, Mock J~J, Justice B~J, Cummer S~A, Pendry J~B, Starr A~F and Smith
  D~R 2006 {\em Science\/} {\bf 314} 977--980

\bibitem{Magnus:2008fw}
Magnus F, Wood B, Moore J, Morrison K, Perkins G, Fyson J, Wiltshire M~C~K,
  Caplin D, Cohen L~F and Pendry J~B 2008 {\em Nat. Mater.\/} {\bf 7} 295--297

\bibitem{Narayana:2011ht}
Narayana S and Sato Y 2011 {\em Adv. Mater.\/} {\bf 24} 71--74

\bibitem{Gomory:2012bo}
G{\"o}m{\"o}ry F, Solovyov M, Souc J, Navau C, Prat-Camps J and Sanchez A 2012
  {\em Science\/} {\bf 335} 1466--1468

\bibitem{Chen:1999uv}
Chen W, Rylyakov A~V, Patel V, Lukens J~E and Likharev K~K 1999 {\em IEEE
  Trans. Appl. Supercond.\/} {\bf 9} 3212--3215

\bibitem{Goldacker:2006fc}
Goldacker W, Nast R, Kotzyba G, Schlachter S~I, Frank A, Ringsdorf B, Schmidt C
  and Komarek P 2006 {\em J. Phys.: Conf. Ser.\/} {\bf 43} 901--904

\bibitem{Nii:2012gf}
Nii M, Amemiya N and Nakamura T 2012 {\em Supercond. Sci. Technol.\/} {\bf 25}
  095011

\bibitem{Bean:1962vg}
Bean C 1962 {\em Phys. Rev. Lett.\/} {\bf 8} 250--253

\bibitem{Swartz:1968dk}
Swartz P~S 1968 {\em J. Appl. Phys.\/} {\bf 39} 4991--4998

\bibitem{Duran:1992wu}
Duran C~A, Gammel P~L, Wolfe R, Fratello V~J, Bishop D~J, Rice J~P and Ginsberg
  D~M 1992 {\em Nature\/} {\bf 357} 474--477

\bibitem{Bellis:1994vb}
Bellis R~H and Iwasa Y 1994 {\em Cryogenics\/} {\bf 34} 129--144

\bibitem{Olson:1997tk}
Olson C~J, Reichhardt C, Groth J, Field S~B and Nori F 1997 {\em Physica C\/}
  {\bf 290} 89--97

\bibitem{Denisov:2006cs}
Denisov D~V, Rakhmanov A~L, Shantsev D~V, Galperin Y~M and Johansen T~H 2006
  {\em Phys. Rev. B\/} {\bf 73} 014512

\bibitem{Colauto:2009gf}
Colauto F, Pati{\~n}o E~J, Aprilli M and Ortiz W~A 2009 {\em J. Phys.: Conf.
  Ser.\/} {\bf 150} 052038

\bibitem{Mawatari:1996we}
Mawatari Y 1996 {\em Phys. Rev. B\/} {\bf 54} 13215--13221

\bibitem{Mawatari:2012hj}
Mawatari Y, Navau C and Sanchez A 2012 {\em Phys. Rev. B\/} {\bf 85} 134524

\bibitem{Nagasawa:2004de}
Nagasawa S, Hinode K, Satoh T, Akaike H, Kitagawa Y and Hidaka M 2004 {\em
  Physica C\/} {\bf 412-414} 1429--1436

\bibitem{Satoh:2005fs}
Satoh T, Hinode K, Akaike H, Nagasawa S, Kitagawa Y and Hidaka M 2005 {\em IEEE
  Trans. Appl. Supercond.\/} {\bf 15} 78--81

\bibitem{Polyanskii:1989vl}
Polyanskii A~A, Vlasko-Vlasov V~K, Indenbom M~V and Nikitenko V~I 1989 {\em
  Sov. Tech. Phys. Lett.\/} {\bf 15} 872

\bibitem{Dorosinskii:1992tp}
Dorosinskii L~A, Indenbom M~V, Nikitenko V~I, Ossip'yan Y~A, Polyanskii A~A and
  Vlasko-Vlasov V~K 1992 {\em Physica C\/} {\bf 203} 149--156

\bibitem{Soibel:2000vf}
Soibel A, Zeldov E, Rappaport M, Myasoedov Y, Tamegai T, Ooi S, Konczykowski M
  and Geshkenbein V 2000 {\em Nature\/} {\bf 406} 282--287

\bibitem{Goa:2001vz}
Goa P~E, Hauglin H, Baziljevich M, Il'yashenko E, Gammel P~L and Johansen T~H
  2001 {\em Supercond. Sci. Technol.\/} {\bf 14} 729--731

\bibitem{Yasugaki:2002hk}
Yasugaki M, Itaka K, Tokunaga M, Kameda N and Tamegai T 2002 {\em Phys. Rev.
  B\/} {\bf 65} 212502

\bibitem{Tsuchiya:2010fi}
Tsuchiya Y, Nakajima Y and Tamegai T 2010 {\em Physica C\/} {\bf 470}
  1123--1125

\bibitem{Altshuler:2004vt}
Altshuler E and Johansen T 2004 {\em Rev. Mod. Phys.\/} {\bf 76} 471--787

\bibitem{Vestgarden:2013is}
Vestgarden J~I, Shantsev D~V, Galperin Y~M and Johansen T~H 2013 {\em
  Supercond. Sci. Technol.\/} {\bf 26} 055012

\bibitem{Prozorov:2006jh}
Prozorov R, Shantsev D and Mints R 2006 {\em Phys. Rev. B\/} {\bf 74} 220511

\bibitem{Altshuler:2004ww}
Altshuler E, Johansen T~H, Paltiel Y, Jin P, Bassler K~E, Ramos O, Chen Q~Y,
  Reiter G~F, Zeldov E and Chu C~W 2004 {\em Phys. Rev. B\/} {\bf 70} 140505

\bibitem{Denisov:2006ho}
Denisov D, Shantsev D, Galperin Y, Choi E~M, Lee H~S, Lee S~I, Bobyl A, Goa P,
  Olsen A and Johansen T 2006 {\em Phys. Rev. Lett.\/} {\bf 97} 077002

\bibitem{Leiderer:1993us}
Leiderer P, Boneberg J, Br{\"u}ll P, Bujok V and Herminghaus S 1993 {\em Phys.
  Rev. Lett.\/} {\bf 71} 2646--2649

\bibitem{Yurchenko:2007fv}
Yurchenko V~V, Shantsev D~V, Johansen T~H, Nevala M~R, Maasilta I~J, Senapati K
  and Budhani R~C 2007 {\em Phys. Rev. B\/} {\bf 76} 092504

\bibitem{Colauto:2008db}
Colauto F, Pati{\~n}o E~J, Blamire M~G and Ortiz W~A 2008 {\em Supercond. Sci.
  Technol.\/} {\bf 21} 045018

\bibitem{Nowak:1997uc}
Nowak E~R, Taylor O~W, Liu L, Jaeger H~M and Selinder T~I 1997 {\em Phys. Rev.
  B\/} {\bf 55} 11702

\bibitem{Behnia:2000uk}
Behnia K, Capan C, Mailly D and Etienne B 2000 {\em Phys. Rev. B\/} {\bf 61}
  3815--3818

\bibitem{Johansen:2002vs}
Johansen T~H, Baziljevich M, Shantsev D~V, Goa P~E, Kang W~N, Kim H~J, Choi
  E~M, Kim M~S and Lee S~I 2002 {\em Europhys. Lett.\/} {\bf 59} 599

\bibitem{Gubin:2005id}
Gubin A~I, Vitusevich S~A, Siegel M and Klein N 2005 {\em Phys. Rev. B\/} {\bf
  72} 064503

\bibitem{Sherrill:1975wa}
Sherrill M~D and Lindstrom W~A 1975 {\em Phys. Rev. B\/} {\bf 11} 1125--1130

\bibitem{Campbell:2007jj}
Campbell A~M 2007 {\em Supercond. Sci. Technol.\/} {\bf 20} 292--295

\end{thebibliography}

\end{document}